\documentclass[pdftex,twocolumn,epjc3,a4paper]{svjour3}

\RequirePackage[T1]{fontenc}
\smartqed  % flush right qed marks, e.g. at end of proof
\usepackage{amsmath}
\usepackage{amssymb}
\RequirePackage{mathptmx}      % use Times fonts if available on your TeX system
\RequirePackage{flushend}
\usepackage{units}
\usepackage{graphicx}
\usepackage{cite}
\usepackage{xcolor} %needed to color text 
\RequirePackage[colorlinks,citecolor=blue,urlcolor=blue,linkcolor=blue]{hyperref}

\newcommand{\CaW}{CaWO$_4$}
\newcommand{\Pbten}{$^{210}$Pb}
\newcommand{\PbTen}{$^{210}$Pb~}

\newcommand{\GeV}{GeV/$c^2$}
\newcommand{\TeV}{TeV/$c^2$}

\journalname{Eur. Phys. J. C}

\begin{document}

\author
{%
G.~Angloher \thanksref{addrMPI}\and
A.~Bento \thanksref{addrCoimbra}\and
C.~Bucci \thanksref{addrLNGS}\and
L.~Canonica \thanksref{addrLNGS}\and
X.~Defay \thanksref{addrTUM}\and
A.~Erb \thanksref{addrTUM,addrWMI}\and
F.~v.~Feilitzsch \thanksref{addrTUM}\and
N.~Ferreiro Iachellini \thanksref{addrMPI}\and
P.~Gorla \thanksref{addrLNGS}\and
A.~G\"utlein \thanksref{addrVienna}\and
D.~Hauff \thanksref{addrMPI}\and
J.~Jochum \thanksref{addrTUE}\and
M.~Kiefer \thanksref{addrMPI}\and
H.~Kluck \thanksref{addrVienna}\and
H.~Kraus \thanksref{addrOxford}\and
J.~C. Lanfranchi \thanksref{addrTUM}\and
J.~Loebell \thanksref{addrTUE}\and
A.~M\"unster \thanksref{addrTUM}\and
C.~Pagliarone \thanksref{addrLNGS}\and
F.~Petricca \thanksref{addrMPI}\and
W.~Potzel \thanksref{addrTUM}\and
F.~Pr\"obst \thanksref{addrMPI}\and
F.~Reindl \thanksref{e1,addrMPI}\and
K.~Sch\"affner \thanksref{addrLNGS}\and
J.~Schieck \thanksref{addrVienna}\and
%S.~Scholl %/\thanksref{addrTUE}\and
S.~Sch\"onert \thanksref{addrTUM}\and
W.~Seidel \thanksref{addrMPI}\and
L.~Stodolsky \thanksref{addrMPI}\and
C.~Strandhagen \thanksref{e2,addrTUE}\and
R.~Strauss \thanksref{addrMPI}\and
A.~Tanzke \thanksref{addrMPI}\and
H.H.~Trinh~Thi \thanksref{addrTUM}\and
C.~T\"urko$\breve{\text{g}}$lu \thanksref{addrVienna}\and
M.~Uffinger \thanksref{addrTUE}\and
A.~Ulrich \thanksref{addrTUM}\and
I.~Usherov \thanksref{addrTUE}\and
S.~Wawoczny \thanksref{addrTUM}\and
M.~Willers \thanksref{addrTUM}\and
M.~W\"ustrich \thanksref{addrMPI}\and
A.~Z\"oller \thanksref{e3,addrTUM}
}
\institute
{%
Max-Planck-Institut f\"ur Physik, D-80805 M\"unchen, Germany \label{addrMPI} \and
Departamento de Fisica, Universidade de Coimbra, P3004 516 Coimbra, Portugal \label{addrCoimbra} \and
INFN, Laboratori Nazionali del Gran Sasso, I-67010 Assergi, Italy \label{addrLNGS} \and
Physik-Department and Excellence Cluster Universe, Technische Universit\"at M\"unchen, D-85748 Garching, Germany \label{addrTUM} \and
Walther-Mei\ss ner-Institut f\"ur Tieftemperaturforschung, D-85748 Garching, Germany \label{addrWMI} \and 
Institut f\"ur Hochenergiephysik der \"Osterreichischen Akademie der Wissenschaften, A-1050 Wien, Austria\\
and Atominstitut, Vienna University of Technology, A-1020 Wien, Austria \label{addrVienna} \and
Eberhard-Karls-Universit\"at T\"ubingen, D-72076 T\"ubingen, Germany \label{addrTUE} \and
Department of Physics, University of Oxford, Oxford OX1 3RH, United Kingdom \label{addrOxford}
}
\thankstext{e1}{corresponding author: florian.reindl@mpp.mpg.de}
\thankstext{e2}{corresponding author: strandhagen@pit.physik.uni-tuebingen.de}
\thankstext{e3}{corresponding author: andreas.zoeller@ph.tum.de}

\title{Results on light dark matter particles with a low-threshold CRESST-II detector}

\date{September 04, 2015}

\maketitle
\begin{abstract}
The CRESST-II experiment uses cryogenic detectors to search for nuclear recoil events induced by the elastic scattering of dark matter particles in \CaW~crystals. Given the low energy threshold of our detectors in combination with light target nuclei, low mass dark matter particles can be probed with high sensitivity. In this letter we present the results from data of a single detector module corresponding to 52 kg live days. A blind analysis is carried out. With an energy threshold for nuclear recoils of \unit[307]{eV} we substantially enhance the sensitivity for light dark matter. Thereby, we extend the reach of direct dark matter experiments to the sub-\GeV~region and demonstrate that the energy threshold is the key parameter in the search for low mass dark matter particles.
\end{abstract}

\section{Introduction}

Today, there is overwhelming evidence for the existence of dark matter. Precision measurements, e.g. of the cosmic microwave background, ascribe roughly one fourth of the energy density of the universe to dark matter, five times more than ordinary matter \cite{Planck2014}. 

However, physicists did not yet succeed to reveal its underlying nature. One of the most favored solutions is the existence of weakly interacting massive particles (WIMPs) thermally produced in the early universe. Those particles are beyond the Standard Model of particle physics and would have a mass in the range of \unit[$\sim$~10]{\GeV} to \unit[1]{\TeV} and an interaction cross section of the weak scale. The Lee-Wein\-berg bound excludes WIMPs lighter than \unit[$\sim$~2]{\GeV}, since they would lead to an overclosure of the universe \cite{LeeWeinbergLimit}. However, in the last years light dark matter particles below the WIMP scale gained more and more interest. In particular, asymmetric dark matter models are considered a viable option as they do not only provide an explanation for dark matter, but also connect its generation to the baryon asymmetry \cite{ReviewAsymmetricDM}.

The Cryogenic Rare Event Search with Superconducting Thermometers -- CRESST -- is one of numerous direct searches around the globe aiming to detect dark matter particles elastically and coherently scattering off target nuclei. To shield against cosmic radiation the experiment is located in the LNGS (Laboratori Nazionali del Gran Sasso) underground laboratory in central Italy. In CRESST-II scintillating \CaW~single crystals are cooled to mK temperatures \cite{Angloher2005_cresstIIproof,Angloher2009_run30}. While most of the energy deposited in a particle interaction induces a phonon (heat) signal, a small fraction (\unit[$\lesssim$ 5]{\%}) is emitted as scintillation light. The phonon signal yields a precise energy measurement, whereas the simultaneously recorded light signal allows for particle identification, a powerful tool to discriminate electron recoils (dominant background) from the sought-for nuclear recoils. The signals of phonon and light detector are measured with independent transition edge sensors (TES) read out by SQUIDs \cite{Angloher2005_cresstIIproof,Angloher2009_run30}. We call the ensemble of phonon and corresponding light detector a detector module. Each detector is equipped with a heater to stabilize the temperature in its operating point in the transition between normal and superconducting state. Via the heater we also inject pulses which are used for the energy calibration and for the determination of the energy threshold.

This technology yields accurately measurable sub-keV energy thresholds and a high-precision energy reconstruction. Combining these features with the light target nuclei, CRESST-II detectors have a unique potential to explore the low-mass regime with dark matter particle masses of below \unit[1]{\GeV}.

\section{CRESST-II Phase 2}

In August 2015 CRESST-II finished its second extensive data taking campaign, referred to as phase 2, featuring 18 detector modules with a total mass of \unit[5]{kg}. Most of the data acquired in two years of measurement time is dark matter data accompanied by calibrations with \unit[122]{keV} $\gamma$-rays ($^{57}$Co-source), high-energy $\gamma$-rays ($^{232}$Th-source) and neutrons (AmBe-source). In 2014 we reported on first results from phase 2, analyzing the detector module with the best overall performance in terms of background level, trigger threshold and background rejection \cite{angloher_results_2014}. This non-blind analysis proved for the first time that CRESST-II detectors provide reliable data for recoil energies down to the threshold of \unit[0.6]{keV} \cite{angloher_results_2014}. Afterwards, we lowered the trigger thresholds of several detectors, achieving \unit[0.4]{keV} for the module used in \cite{angloher_results_2014} and \unit[0.3]{keV} for the module Lise, which is the lowest value obtained in phase 2.

In contrast to the module used for the 2014 result \cite{angloher_results_2014}, Lise is not equipped with the upgraded crystal holding design which uses \CaW-sticks instead of metal clamps. The upgraded holding scheme avoids any non-scintillating surface in line-of-sight to the crystal, thus efficiently vetoing backgrounds induced by $\alpha$-decays on or slightly below surfaces \cite{strauss_detector_2015}. We want to emphasize that the lower threshold of Lise is not connected to the holding concept, but solely arising from a superior performance of the phonon detector. 

Based on the method already discussed in \cite{angloher_results_2014} we analyze \unit[52.2]{kg days} of data taken with the module Lise (\unit[300]{g}) and its threshold set at \unit[0.3]{keV}. A blind analysis is carried out by first defining a statistically insignificant part of the data set as a training set. In the previous phase 1 \cite{angloher_results_2012} we proved that CRESST-II detectors can be operated stably over long time periods, in particular concerning the energy reconstruction. This is also confirmed for Lise as will be discussed in section~\ref{sec:LongTerm}. Thus, the training set is well suited to develop all methods of data preparation and selection. We then apply them blindly -- without any change -- to the final data set. Data from the training set is not part of the final analysis.

For cross checks and validation four independent analyses covering the complete process from raw data to final result are performed. Based on the training set results one analysis is chosen a-priori. However, the results on training set and final data of all different analyses are consistent.

\section{Energy Threshold and Signal Survival Probability}

\subsection{Energy Threshold}

\begin{figure}[tbh]
  \centering
  \includegraphics[width=\columnwidth]{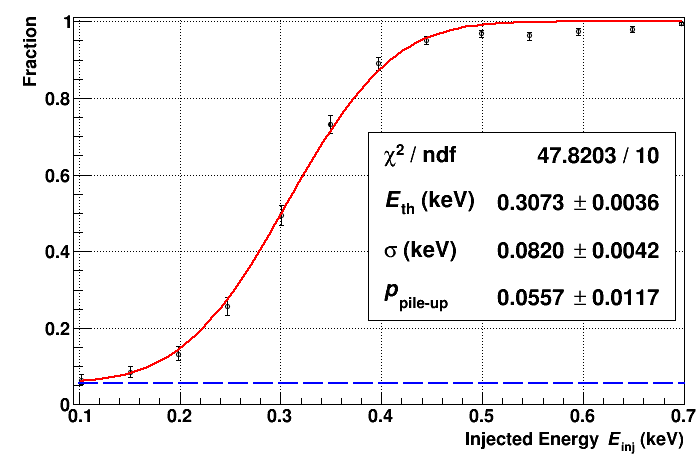}
  \caption{Fraction of heater pulses triggering, injected with discrete energies $E_{\text{inj}}$. The error bars indicate the statistical (binomial) uncertainty at the respective energy. 
  The solid red curve is a fit with the sum of a scaled error function and a constant pile-up probability $p_{\text{pile-up}}$ (blue dashed line). 
  The latter is independent of the injected energy since it is caused by random coincidences of the injected heater pulses with any other trigger.
  The fit yields an energy threshold of $E_{\text{th}}=\unit[(307.3\pm 3.6)]{eV}$ and a width of \unit[$\sigma=(82.0\pm 4.2)$]{eV}.}
  \label{fig:TriggerEfficiencies}
\end{figure}

In the light of the expected exponential rise of the dark matter particle spectrum towards lower recoil energies a precise knowledge of the energy threshold is mandatory. The finite baseline noise causes a deviation of the trigger efficiency from an ideal step function. We directly measure the trigger efficiency by injecting low-energy heater pulses with a shape similar to pulses induced by particle interactions.
For each injected energy $E_{\text{inj}}$ the fraction of pulses causing a trigger is determined as depicted in figure~\ref{fig:TriggerEfficiencies}.

The injected energy $E_{\text{inj}}$ is calibrated with \unit[122]{keV} $\gamma$-rays and is verified with low-energy background lines (see section~\ref{sec:energyCalibration}).
The solid curve (red) is a fit with the function
\begin{equation}
f(E_{\text{inj}}) = \frac{1-p_{\text{pile-up}}}{2} \cdot \left[ 1 + \text{erf} \left(
\frac{E_{\text{inj}} - E_{\text{th}}}{ \sqrt{2}\sigma} \right) \right] +
p_{\text{pile-up}}
\end{equation}
where erf denotes the Gaussian error function; 
$f(E_{\text{inj}})$ describes the probability for a heater pulse with injected energy $E_{\text{inj}}$ to trigger. 
The pile-up probability $p_{\text{pile-up}}$ (blue dashed line) accounts for random coincidences of the injected heater pulse with any other trigger and is, therefore, independent of the injected energy $E_{\text{inj}}$. 
The resulting value of $p_{\text{pile-up}} =$ \unit[(5.57 $\pm$ 1.17)]{\%} is compatible with the expectation based on the trigger rate in this module and the respective coincidence time window. The error function describes a convolution of a step function for an ideal threshold $E_{\text{th}}$ with the baseline noise modelled by a Gaussian. The fit yields an energy threshold of $E_{\text{th}} = \unit[(307.3\pm 3.6)]{eV}$ and a width of $\sigma = \unit[(82.0\pm 4.2)]{eV}$.
The latter is in good agreement with the widths of low-energy $\gamma$-lines (see table~\ref{tab:energies}) confirming that the resolution at low energies is dominated by baseline noise.

\subsection{Signal Survival Probability}

For all cuts applied to the data, energy dependent signal survival probabilities are determined by performing the cuts on a set of artificial pulses with discrete energies. 
These are created by superimposing signal templates and empty baselines, i.e. random triggers periodically sampled throughout the whole data taking period. We obtain the signal templates by averaging pulses of equal energy. For this analysis $\gamma$-events in the \unit[122]{keV} $^{57}$Co-calibration peak are used. 
The templates of phonon and light detector are scaled respectively to achieve the desired energy. 
A dedicated neutron calibration confirms that pulse shape differences between electron and nuclear recoils are negligible. 
Therefore, by accounting for the reduced light output of nuclear recoil events (quenching) artificial signal events are created. 

\begin{figure}[htb]
  \centering
  \includegraphics[width=\columnwidth]{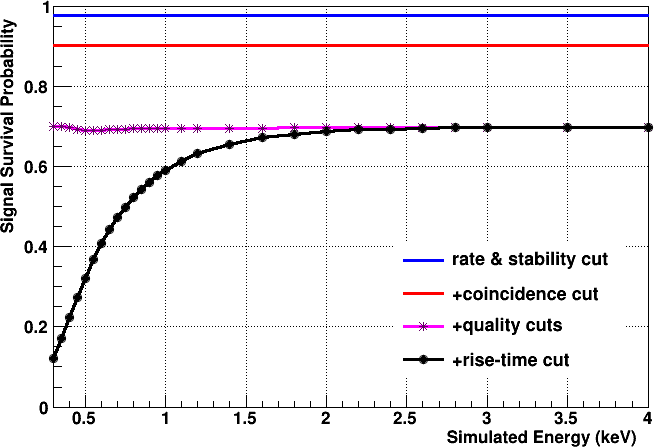}
  \caption{The solid lines represent the signal survival probability after successive application of the cuts, as discussed in the text. The simulated pulses correspond to nuclear recoil events at discrete energies starting from the threshold of \unit[0.3]{keV} (data points).}
  \label{fig:Efficiencies}
\end{figure}

The fraction of signals with a certain simulated energy passing each cut yields the respective survival probability. In figure~\ref{fig:Efficiencies} we show the cumulative probability after each selection criterion which are discussed in the following.

The first cut removes time periods where the trigger rate exceeds a pre-defined limit.
Such periods are mainly caused by an enhanced baseline noise contribution from mechanical vibrations induced by the cryogenic facility.
For technical reasons the \textit{rate cut} is performed together with the \textit{stability cut}, which only accepts time periods with both detectors of a module stably running at their corresponding operating points.
Rate and stability cut in total remove \unit[2.32]{\%} of the collected exposure.

Due to the low interaction probability multiple scatterings are not expected for dark matter particles. 
Therefore, all events coincident with a signal in any other detector module and/or the muon veto are rejected by the \textit{coincidence cut}, reducing the exposure by another \unit[7.72]{\%}.

 We reconstruct the energy by fitting the signal template to the measured pulse of the corresponding detector and quantify the fit quality by the Root Mean Square (RMS) deviation. Events where a correct energy reconstruction is not guaranteed are mainly removed by a cut on the RMS. The line labeled \textit{quality cuts} in figure~\ref{fig:Efficiencies} shows the combined impact of the RMS cut and additional cuts addressing specific artifacts (e.g. pile-up events and SQUID resets). 

For Lise the TES of the phonon detector is evaporated onto a small \CaW~crystal, called TES-carrier, which is in turn glued onto the main absorber crystal (composite design, details can be found in \cite{Kiefer2009}). 
Due to the different thermal couplings, pulses in the carrier exhibit a different pulse shape compared to those in the main absorber, mainly visible in faster rise and decay times.
In Lise only a few TES-carrier-like events were observed in the training set. Nonetheless, a strict cut on the \textit{rise-time parameter}  was applied to remove carrier events with a very high efficiency.
This has some cost on the signal survival probability.
One of the cross-check analyses uses an artificial neural network to tag the TES-carrier like events, exploiting further pulse shape parameters.
For Lise no significant impact on the result is observed, confirming the performance of the cut on the rise time.
In all cases a non-negligible fraction of potential signal events, especially at low energies (\unit[< 2]{keV} in figure~\ref{fig:Efficiencies}), is removed. 
Therefore, for CRESST-III\footnote{Due to the major detector upgrades foreseen, the experiment will be renamed as CRESST-III.} we will directly evaporate the TES on the absorber crystal, avoiding the TES-carrier.

For Lise we determine a final signal survival probability at threshold of \unit[12]{\%} which is roughly two times higher compared to the 2014 result \cite{angloher_results_2014}.

\section{Energy Calibration and Discussion}
\label{sec:energyCalibration}

The energy calibration uses heater pulses of several defined energies which are injected periodically throughout the entire data taking period \cite{Angloher2005_cresstIIproof,Angloher2009_run30}. These pulses are of similar shape compared to those induced by particle interactions. The heater pulses are used to define a linear response over the entire dynamic range of the detector. The absolute energy scale is finally set by evaluating \unit[122]{keV} $\gamma$-rays during dedicated calibrations with a $\mathrm{^{57}}$Co-source. 

\begin{figure}[htbp]
  \includegraphics[width=\columnwidth]{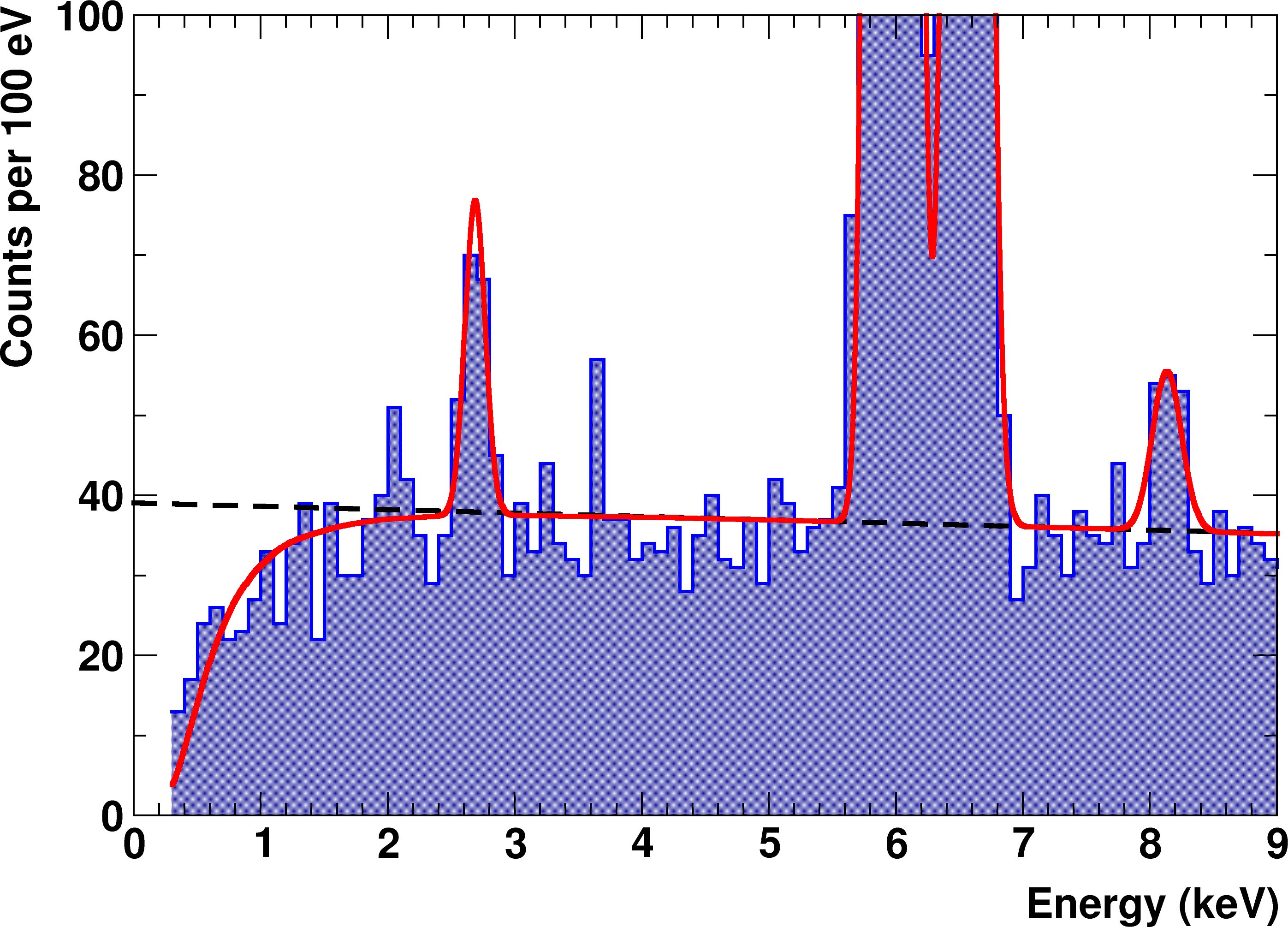}
  \caption{Spectrum of all events (blue) up to \unit[9]{keV}, truncated at a bin content of 100. The red line corresponds to a data-driven background model including a linearly decreasing background (dashed black line) and contributions from $^{179}$Ta (M1) (\unit[2.7]{keV}), $^{55}$Fe (\unit[6.0]{keV} \& \unit[6.6]{keV}), and Cu fluorescence (\unit[8.1]{keV}). The drop towards lower energies results from the declining signal survival probability (see figure \ref{fig:Efficiencies}). This plot illustrates that the detector Lise exhibits an almost flat background level down to the threshold energy. 
}
  \label{fig:EnergySpectrumModel}
\end{figure}

The energy spectrum after cuts is shown in figure~\ref{fig:EnergySpectrumModel}. The peak at \unit[2.7]{keV} is due to cosmogenic activation of tungsten, the copper fluorescence line can be seen at \unit[8.1]{keV}. The dominant double peak around \unit[6]{keV} originates from an accidental illumination with an $^{55}$Fe X-ray source installed to calibrate the light detector of a close-by detector module. Despite the significant count rate in the double peak, it only marginally impacts the sensitivity for low-mass dark matter particles because of its high energy and distinct line shape.

Depicted in figure~\ref{fig:EnergySpectrumModel} is the result from a fit of a data-driven background model. The model includes the signal survival probability, as discussed previously, a linearly decreasing background (dashed black line) and the observed low-energy lines. The positions of the peaks are free parameters in the unbinned likelihood fit of the model, whereas the widths of the peaks are described by a linear function allowing for an energy dependence of the resolution. 

As mentioned earlier, the artificial nuclear recoil events, as used to determine the signal survival probability, are created by superimposing baseline noise samples with signal templates. For those artificial events the baseline noise -- equivalent to the resolution at zero energy -- is the only effect causing a finite resolution of the reconstructed energy. This method provides a more precise and -- due to the high number of empty baselines -- statistically superior estimate compared to a fit of lines in the low-energy spectrum. The determined value of $\sigma_0=$\unit[$(62\pm1)$]{eV} is included as a fixed parameter in the background model, whereas the linear term may be varied by the fit. 

A comparison between literature values $\mathrm{E_{lit}}$ and fitted energies $\mathrm{E_{fit}}$ of low-energy lines can be found in table~\ref{tab:energies}. Although we do not use these lines for calibration we find an excellent agreement with the nominal energies, demonstrating the outstanding accuracy of our energy reconstruction. Listed in the last column of table~\ref{tab:energies} is the energy resolution evaluated at the respective position of the line\footnote{The values obtained are compatible with uncorrelated fits of the lines, not constraining the Gaussian width.}.     

  \begin{table}[htb]
  \centering
  \caption{Comparison between literature values and fitted values for the low-energy lines observed in the spectrum of the module Lise (figure~\ref{fig:EnergySpectrumModel}). Literature values are taken from \cite{firestone} (weighted with intensity if needed). Energy resolutions evaluated at $\mathrm{E_{fit}}$ are listed in addition.}
  
  \label{tab:energies}
  \begin{tabular}{ccccc}
  \hline
 Origin & $\mathrm{E_{lit}}$ {(}keV{)} & $\mathrm{E_{fit}}$ {(}keV{)} & $\frac{\mathrm{E_{fit}}-\mathrm{E_{lit}}}{\mathrm{E_{lit}}}$ (\%) & $\mathrm{\sigma}$ {(}eV{)} \\ \hline
 $^{179}$Ta & 2.601                     & 2.687 $\pm$ 0.020         & +3.3                 & 79  $\pm$ 1          \\
 Mn K$_\alpha$ & 5.895                     & 5.972 $\pm$ 0.002         & +1.3                 & 101 $\pm$ 1         \\
 Mn K$_\beta$ & 6.490                     & 6.562 $\pm$ 0.003         & +1.1                 & 105 $\pm$ 2         \\
 Cu &8.041                     & 8.133 $\pm$ 0.034         & +1.1                 & 115 $\pm$ 2          \\ \hline
  \end{tabular}
  \end{table}   

  For every particle interaction the total energy deposited in the crystal is shared between the phonon and the light detector. This split-up differs for different event classes, in particular for $\gamma$-rays (used for calibration) compared to nuclear recoils (potential dark matter signal). Nuclear recoils produce less light and, thus, more energy remains in the phonon detector. In \cite{angloher_results_2014} we successfully demonstrated that a precise measurement of the scintillation light allows to take into account this effect accurately. Typical light detectors operated in phase 2 (like the one used in \cite{angloher_results_2014}) have an energy resolution which is one order of magnitude better than the one of Lise. For the energies of interest, the light signal in Lise is, therefore, dominated by statistical fluctuations of the baseline noise. Thus, we do not apply the correction as used in \cite{angloher_results_2014}. This causes the slight overestimation of the reconstructed energies in the order of a few percent, as can be seen in column four of table \ref{tab:energies}. In any case, we find this deviation to have negligible influence on the final result.

\section{Long-Term Stability}
\label{sec:LongTerm}

\begin{figure}[htb]
  \includegraphics[width=\columnwidth]{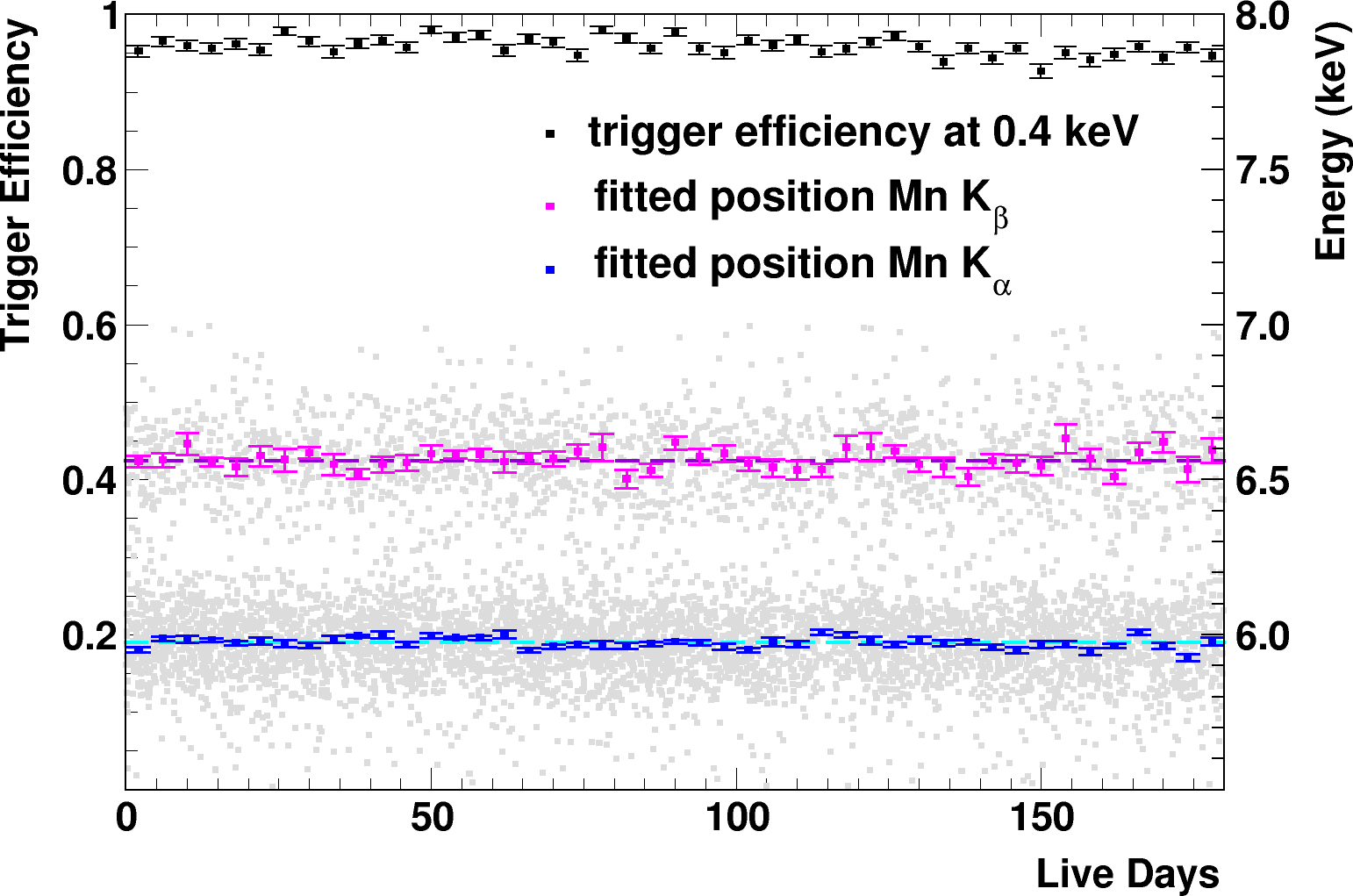}
  \caption{Trigger efficiency (black, left y-axis) as a function of the complete measurement time. Each data point is determined as the fraction of \unit[0.4]{keV} heater pulses triggering in the respective time bin. For the same time bins also the fitted peak positions and the fit errors for Mn K$_{\alpha}$ (blue) and K$_{\beta}$ (magenta) from the $^{55}$Fe-source are drawn (scale on the right y-axis). Only small deviations from the fitted total mean values (dashed lines) are observed.
The constant spread of the corresponding data (grey points) illustrates a stable line width.
}
  \label{fig:LongtermStability}
\end{figure}
  
The sensitivity for light dark matter critically depends on the ability to trigger and reconstruct events of very low energies. 
To prove a stable trigger efficiency over time we periodically inject low-energy heater pulses (corresponding to $\sim$~\unit[0.4]{keV}). In figure~\ref{fig:LongtermStability} the fraction of these pulses triggering is shown in black together with statistical (binomial) errors. 
The trigger efficiency at \unit[0.4]{keV} is slightly higher (\unit[$\sim$95]{\%}) in the final data set compared to the one determined in figure~\ref{fig:TriggerEfficiencies} (\unit[$\sim$90]{\%}). Thus, the value of \unit[307]{eV} measured for the threshold energy is conservative.

The high statistics in the $^{55}$Fe double-peak allows us to monitor the accuracy of our energy reconstruction and the stability of the energy resolution at low energies over time. In each time bin a combined fit of two Gaussians is performed to extract the reconstructed peak positions of the $\mathrm{K_{\alpha}}$ (blue) and $\mathrm{K_{\beta}}$ (magenta) lines (see figure~\ref{fig:LongtermStability}). The small variations of the peak positions show the very stable operation of the detector. In addition, a scatter plot of the relevant events (grey dots) further illustrates that the means and also the line widths of the two peaks remain stable over time.

\section{Light Yield}

\begin{figure}[htb]
  \centering
  \includegraphics[width=\columnwidth]{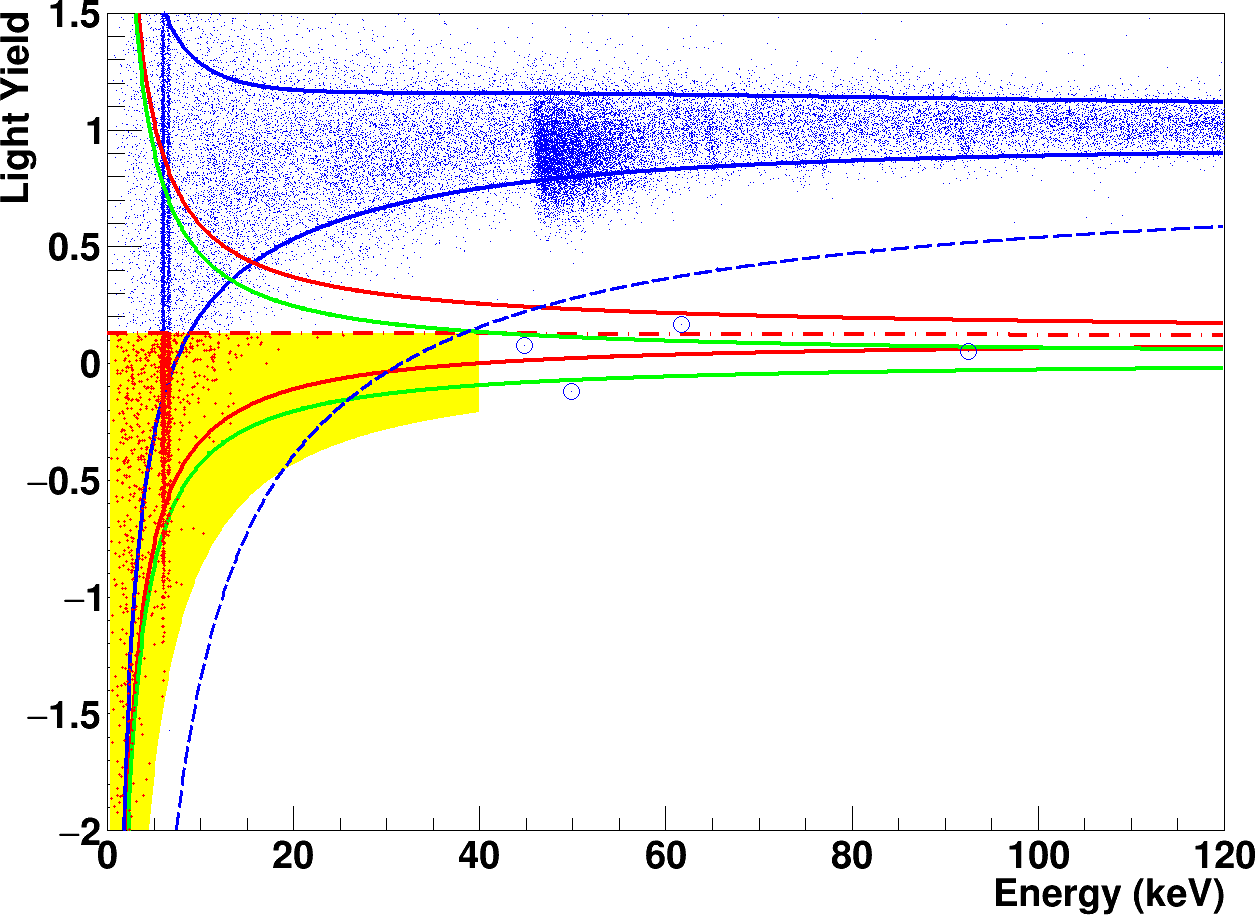}
  \caption{Data taken with the detector module Lise depicted in the light yield - energy plane. The solid lines mark the \unit[90]{\%} upper and lower boundaries of the e$^-/\gamma$-band (blue), the band for recoils off oxygen (red) and off tungsten (green). The dashed blue line corresponds to the lower \unit[5]{$\sigma$} boundary of the e$^-/\gamma$-band, events outside are marked with a blue circle. The upper boundary of the acceptance region (yellow area) is set to the middle of the oxygen band (dashed dotted red), the lower one to the \unit[99.5]{\%} lower boundary of the tungsten band. Events therein are additionally highlighted in red.} 
  \label{fig:LYPlotMarked}
\end{figure}

In figure \ref{fig:LYPlotMarked} we present the light yield -- defined for every event as the ratio of light to phonon signal -- as a function of energy. Recoils on electrons have the highest light output, set to one by calibration (at \unit[122]{keV}). The solid blue lines mark the \unit[90]{\%} upper and lower boundaries of the e$^-/\gamma$-band, with \unit[80]{\%} of electron recoil events expected in-between. The band is determined by an unbinned likelihood fit to the data \cite{strauss_detector_2015}. 
Two prominent features can be seen in the  e$^-/\gamma$-band, the double peak at \unit[$\sim6$]{keV} originating from the $^{55}$Fe-source and a $\beta$-decay spectrum from an intrinsic contamination in the crystal with \PbTen starting at \unit[46.5]{keV}. As mentioned before, in this analysis we do not correct the energy for events exhibiting a different energy sharing between phonon and light detector compared to the \unit[122]{keV} $\gamma$-rays used for calibration. Thus, small statistical fluctuations in light output cause a slight tilt of the left shoulder of the \Pbten~$\beta$-spectrum \cite{angloher_results_2014,Arnaboldi2010}.

Nuclear recoils produce less light than electron recoils, as quantified by the quenching factor for the respective target nucleus. The quenching factors, including their energy dependence, are precisely known from dedicated external measurements \cite{strauss_energy-dependent_2014} and allow to analytically calculate the nuclear recoil bands for scatterings off tungsten (solid green), calcium (not shown) and oxygen (solid red), starting from the e$^-/\gamma$-band. The validity of this approach is, for example, shown in \cite{angloher_results_2012} and was confirmed with a neutron calibration. 

There are four events (blue circles) in the data well below the \unit[5]{$\sigma$} lower boundary of the e$^-/\gamma$-band (dashed blue). These events are statistically incompatible with leakage and, thus, indicate the presence of an additional source of events in this module, apart from e$^-/\gamma$-background. As discussed in the introduction, Lise is of a conventional module design using non-scintillating clamps to hold the crystal and, therefore, is not capable of vetoing all events originating from $\alpha$-decays on or slightly below surfaces \cite{angloher_results_2012,strauss_detector_2015}. Additional data from Lise (before lowering the threshold) and the ten other conventional modules further underpins this explanation. 

\section{Leakage into the Acceptance Region}
\subsection{Acceptance Region}

Before unblinding we define the acceptance region (yellow region in figure \ref{fig:LYPlotMarked}) to extend from the threshold energy of \unit[307]{eV} to \unit[40]{keV}. In light yield the acceptance region spans from the \unit[99.5]{\%} lower boundary of the tungsten band to the center of the oxygen band (dashed-dotted red line in figure \ref{fig:LYPlotMarked}). For the calculation of the exclusion limit all events inside the acceptance region (highlighted in red) are conservatively considered as potential signal events. In \cite{angloher_results_2014} we found that the center of the oxygen band is the optimal choice in terms of limiting the leakage from the e$^-/\gamma$-band on the one hand and sampling the signal of all three nuclear recoil bands on the other hand. This finding was confirmed for Lise based on training set data. However, the impact of the choice of reasonable light yield boundaries is on the level of the expected statistical fluctuations.

\subsection{Background Leakage}

In the energy range from 1 to \unit[40]{keV} we determine an average background level of \mbox{\unit[13]{counts/(keV kg day)}}.
If the contribution of the $^{55}$Fe-source is excluded we obtain a value of \mbox{\unit[8.5]{counts/(keV kg day)}} which is still well above the value of \mbox{\unit[3.51]{counts/(keV kg day)}}\cite{strauss_betagamma_2015} determined for the module used in \cite{angloher_results_2014}. 
These differences almost completely arise from different levels of crystal-intrinsic radioactive contaminations. The fact that the crystal of the module Lise is from an external supplier while the crystal analyzed in 2014 \cite{angloher_results_2014} was grown at Technische Universit\"at M\"unchen further stresses the importance of in-house crystal growth \cite{TUMCrystal}.

\begin{figure}[htb]
  \centering
  \includegraphics[width=\columnwidth]{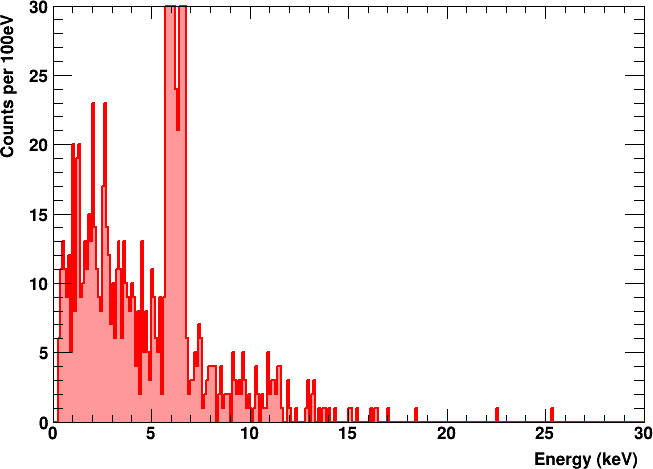}
  \caption{Energy spectrum of all events in the acceptance region (\unit[307]{eV}--\unit[40]{keV}, see figure \ref{fig:LYPlotMarked}) truncated at a bin content of 30 for reasons of clarity. For the final result all events are conservatively considered as potential signal events to extract an exclusion limit using Yellin's optimum interval method.}
  \label{fig:AccEvents}
\end{figure}

Since the widths of the bands are dominated by the energy resolution of the light detector, the modest light detector performance of the module Lise results in a low discrimination power between the sought-for nuclear recoils and the dominant e$^-/\gamma$-background. Thus, we observe leakage of e$^-/\gamma$-events into the acceptance region, leading to the energy spectrum of all accepted events depicted in figure~\ref{fig:AccEvents}.

\section{Expected Signal Composition on the Target Nuclei}
\label{sec:CountsOnTarget}

To calculate the expected recoil spectrum of dark matter particles we make use of the following standard assumptions on the dark matter halo: Maxwellian velocity distribution, asymptotic velocity of \unit[220]{km/s}, galactic escape velocity of \unit[544]{km/s} and local dark matter density of \unit[0.3]{GeV/cm$^3$}. We use the Helm parametrization of the form factors \cite{Lewin1996_reviewmathematics} quantifying effects of the nuclear substructure on the scattering cross-section. Finally, the choice of the acceptance region in the light yield - energy plane and the energy resolution at zero energy (baseline noise) are taken into account. For the latter we use the value of $\sigma_0=$\unit[62]{eV} as determined by the evaluation of the artificial signal events (see section~\ref{sec:energyCalibration}).

\begin{figure}[htb]
  \centering
  \includegraphics[width=\columnwidth]{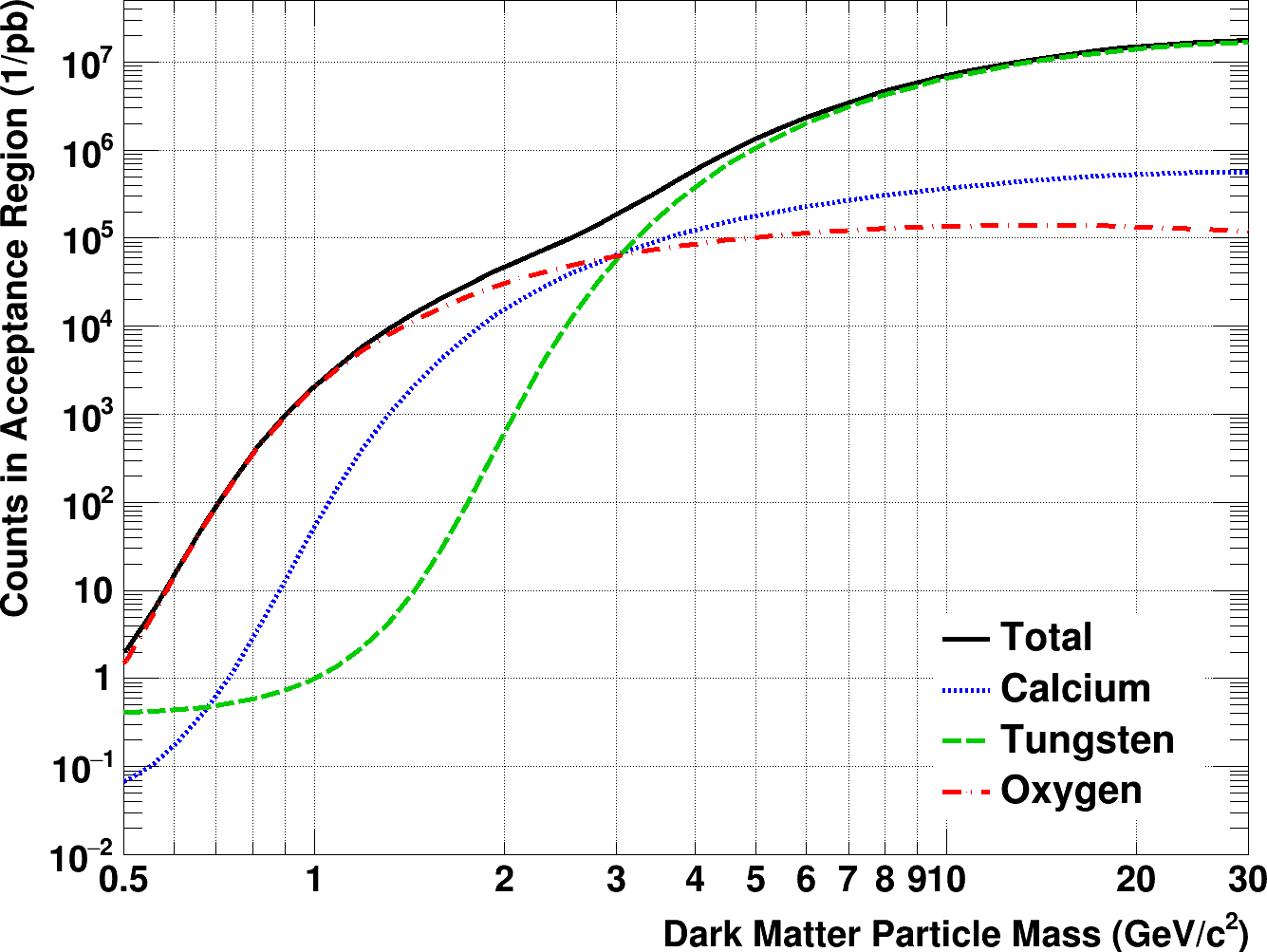}
  \caption{Expected number of counts in the acceptance region for a dark matter-nucleon cross-section of \unit[1]{pb} and standard astrophysical assumptions as a function of the mass of the dark matter particle. The solid black line depicts the overall count number which is the sum of the number of scatterings off the three target nuclei (colored lines, see legend). }
  \label{fig:totRate}
\end{figure}

Figure \ref{fig:totRate} shows the number of expected events to be observed by the detector Lise in the final data set as a function of the mass of the dark matter particle. The lines are normalized to a dark matter particle-nucleon cross section of \unit[1]{pb} (\unit[$\equiv10^{-36}$]{cm$^2$}). The colored curves (see legend) correspond to the three target nuclei oxygen, calcium and tungsten. The total sum of the three constituents is drawn in solid black. 

Due to the anticipated A$^2$-dependence of the cross section, dark matter particles are supposed to dominantly scatter off the heavy tungsten. The energy transferred in the scattering process is a function of the reduced mass of target nucleus and dark matter particle. Thus, for a given mass of the dark matter particle the fraction of the expected energy spectrum above threshold depends on the mass of the target nucleus. 

As a result, for dark matter particles with masses above \unit[5]{\GeV} recoils off tungsten are expected to be far more numerous compared to oxygen and calcium. For lighter masses a substantial part of the tungsten recoils have energies below threshold leading to a strong decrease of the number of counts. This results in a mass range completely dominated by scatterings off oxygen, because the drop for oxygen and calcium is shifted towards lower masses (see figure~\ref{fig:totRate}).

In the limit of very low masses, the reduced mass converges to the mass of the dark matter particles, causing less pronounced differences in the shape of the recoil spectra on the different target nuclei. This effect is further augmented by the influence of the baseline noise. Since the A$^2$-scaling of the cross sections still persists, scatterings off tungsten account for a slightly larger proportion of the total expected signal again.

\section{Result, Discussion and Outlook}

For each dark matter particle mass we use the Yellin optimum interval method \cite{Yellin2002_Limit,optimum_I} to calculate an upper limit with \unit[90]{\%} confidence level on the elastic spin-independent interaction cross-section of dark matter particles with nucleons. While this one-dimensional method does not rely on any assumption on the background, it exploits differences between the measured (see figure \ref{fig:AccEvents}) and the expected energy spectrum (see section \ref{sec:CountsOnTarget}). 

\begin{figure}[htb]
  \includegraphics[width=\columnwidth]{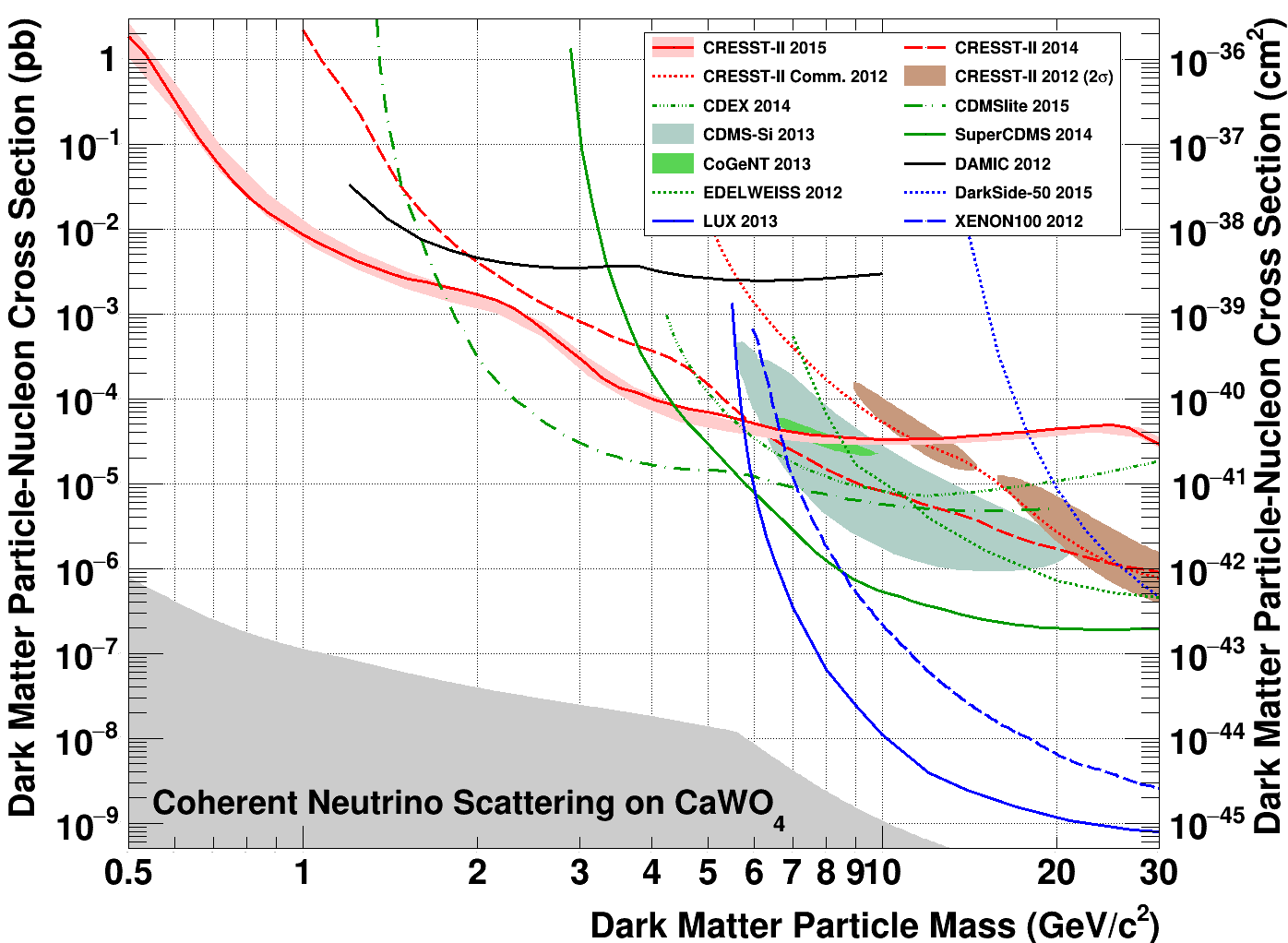}
  \caption{Parameter space for elastic spin-independent dark matter-nucleon scattering. The result from this blind analysis is drawn in solid red together with the expected sensitivity ($1\sigma$ confidence level (C.L.)) from the data-driven background-only model (light red band). The remaining red lines correspond to previous CRESST-II limits \cite{angloher_results_2014,brown_extending_2012}. 
The favored parameter space reported by CRESST-II phase 1 \cite{angloher_results_2012}, CDMS-Si \cite{Agnese2013} and CoGeNT \cite{Aalseth2013} are drawn as shaded regions. 
For comparison, exclusion limits (\unit[90]{\%} C.L.) of the liquid noble gas experiments \cite{agnes_first_2015,Akerib2014,Aprile2012} are depicted in blue, from germanium and silicon based experiments in green and black \cite{cdex_collaboration_limits_2014,supercdms_collaboration_wimp-search_2015,Agnese2014,barreto_direct_2012,edelweiss_collaboration_search_2012}. 
In the gray area coherent neutrino nucleus scattering, dominantly from solar neutrinos, will be an irreducible background for a CaWO$_4$-based dark matter search experiment \cite{gutlein_impact_2015}. }

  \label{fig:LimitPlot}
\end{figure}

The resulting exclusion limit of this blind analysis is drawn in solid red in figure \ref{fig:LimitPlot}. For higher masses this module does not have a competitive sensitivity, due to the large number of background events. In particular, the leakage from the $^{55}$Fe-source (see figure~\ref{fig:AccEvents}) results in an almost flat limit for masses of \unit[5--30]{\GeV}. However, for dark matter particles lighter than \unit[1.7]{\GeV} we explore new regions of parameter space.

The improvement compared to the 2014 result \cite{angloher_results_2014} (red dashed line) is a consequence of the almost constant background level down to the threshold which was reduced from \unit[603]{eV} to \unit[307]{eV}. The lower the mass of the dark matter particle the more relevant these improvements become. With this analysis we explore masses down to \unit[0.5]{\GeV}, a novelty in the field of direct dark matter searches.

The transition point of the dominant scattering target nucleus manifests itself as kink in the corresponding exclusion curve. Due to the lower threshold Lise starts to be dominated by scatterings off tungsten already at \unit[$\sim$3]{\GeV} (see figure~\ref{fig:totRate}) compared to \unit[$\sim$4.5]{\GeV} for the 2014 result \cite{angloher_results_2014}.

Due to the rather large number of leakage events into the acceptance region the result is already not limited by exposure any more. Consequently, only small statistical fluctuations are expected. This is confirmed by calculating limits for 10,000 Monte Carlo sets sampled from the data-driven background model discussed in section~\ref{sec:energyCalibration}. The resulting \unit[1]{$\sigma$} contour is shaded in light red in figure \ref{fig:LimitPlot}.

In CRESST-III we will substantially size down the absorber crystals in order to achieve lower energy thresholds. Furthermore, we expect two beneficial effects on the light signals: Firstly more light reaches the light detector and secondly the light detector can also be scaled down which leads to an enhanced energy resolution. Both improvements will increase the background discrimination power.
All modules will feature an upgraded holding scheme and will mainly be equipped with absorber crystals produced in-house due to their significantly lower level of intrinsic radioactive contaminations.
Combining these measures with the enhanced discrimination power, a drastically reduced background leakage is expected. 

In this letter we prove that a low energy threshold is the key requirement to achieve sensitivity to dark matter particles of $\mathcal{O}($\unit[1]{\GeV}) and below. We expect significant progress exploring the low mass regime with the upcoming CRESST-III experiment, featuring next-generation detectors optimized towards the detection of recoil energies as small as \unit[100]{eV}.

\begin{acknowledgements}
We are grateful to LNGS for their generous support of CRESST, in particular to Marco Guetti for his constant assistance.
This work was supported by the DFG cluster of excellence: Origin and Structure of the Universe, by the Helmholtz Alliance for Astroparticle Physics, and by the BMBF: Project 05A11WOC EURECA-XENON.
\end{acknowledgements}

\bibliographystyle{h-physrev}
\bibliography{biblio.bib}
\end{document}